\documentstyle[prl,aps,multicol]{revtex}
\tighten
\begin{document}
\draft
\title{
Mesoscopic charge fluctuations in the Coulomb blockade regime
}
\author{A. Kaminski$^{1}$, I.L. Aleiner$^{2}$, and L.I. Glazman$^{1}$}
\address{$^{1}$Theoretical Physics Institute, University of
Minnesota, Minneapolis, MN 55455\\
$^{2}$Department of Physics, State University of New York, Stony Brook, NY 11794}
\maketitle

\begin{abstract} 
We study mesoscopic fluctuations of the differential capacitance of a dot 
connected to a lead by a single-mode quantum channel with adjustable conductance 
$G$. We show that the amplitude of the fluctuations reaches maximum at a 
partially opened channel, $G\lesssim e^2/\pi\hbar$. Parametric correlations of 
fluctuations at different values of the applied voltage and magnetic field can 
be studied experimentally, and we find the corresponding correlation functions 
for the differential capacitance. 
\end{abstract} 

\begin{multicols}{2} 
Charge of a conducting dot weakly connected to leads is 
quantized in the units of electron charge $e$. When the potential induced on the 
dot is varied, its charge varies in a step-like manner. (The potential can be 
tuned by means of voltage $V_g$ applied to a gate, which is coupled to the dot 
only electrostatically.) The step-like behavior of the dot charge $Q(V_g)$, {\em 
i.e.}, Coulomb blockade, was clearly demonstrated in experiments utilizing 
metallic islands\cite{Devoret}.  The charge quantization is progressively 
smeared with the increase of the conductance $G_0$ of a junction connecting dot 
with a lead. Theory\cite{GlazmanMatveev} predicts that deviations of $Q$ from 
the quantized values are proportional to conductance, $|Q(V_g)-me|\propto G_0$ 
at $G_0\ll G_q$.  When $G_0$ is of the order of the quantum unit $G_q\equiv 
e^2/\pi\hbar$,  the function $Q(V_g)$ depends on the detailed properties of the 
junction \cite{Schoen,Flensberg,Matveev}. For a single-mode junction, the 
Coulomb blockade vanishes when  the reflectionless propagation of an electron 
mode through the  dot--lead channel is almost reached,  with the periodic part 
of $Q(V_g)$ approaching zero approximately as $G_q-G_0$, see\cite{Matveev}. 
Single-mode junction allowing for ballistic propagation of electrons is a 
reasonable model for a semiconductor quantum dot device. In such a device, $G_0$ 
is tunable, unlike in the lithographically prepared metallic 
samples\cite{Devoret}. This allowed several groups to attack the challenging 
problem of quantum charge fluctuations in the Coulomb blockade regime. 
Suppression of the Coulomb blockade effect at $G_0\to G_q$ was demonstrated 
experimentally by Waugh {\em et al.}\cite{Westervelt} and Molenkamp {\em et 
al.}\cite{Molenkamp} on double-dot devices.  
In very recent experiments, Berman                                
{\em et al.}\cite{Zhitenev} measured the charge smearing in 
a single dot.

Mesoscopic fluctuations of the charge $Q$ and correspondingly of the
differential capacitance $C_{\it diff}$ of a dot reflect the randomness of the
electron states within the dot. Gopar {\em et. al.}\cite{Buttiker} employed the Random
Phase Approximation to consider the charge
dynamics of the dot connected to a reservoir by a reflectionless ($G_0=G_q$)
one mode channel. In this approach, the mesoscopic fluctuations are
associated with the fluctuations of the density of states on the Fermi
level. Later it was shown\cite{AleinerGlazman97} that even in this regime the Coulomb
interaction leads to the formation of a collective state involving
one-electron states in a wide energy strip around the Fermi level
(width of the strip is of the order of the charging energy $E_C$).
As the result, the mesoscopic fluctuations of the capacitance at $G_0=G_q$ are larger
than predicted in Ref.\cite{Buttiker}, and the scale of
the correlation magnetic field is controlled by the charging energy
rather than by one electron level spacing.

In this Letter, we will study the capacitance fluctuations in the whole region 
$0<G_0<G_q$. We will show that the mesoscopic fluctuations are determined by the 
fluctuations of a large number ($\simeq E_C/\Delta$, where $\Delta$ is the 
one-electron level spacing) of single electron wave functions rather than by 
the state at the Fermi level. Concentrating on the low temperature limit 
$T \lesssim \Delta$, we show that the dependence of the variance $\langle\delta 
C^2_{\it diff}\rangle$ on conductance $G_0$ is non-monotonous, with the maximum of 
fluctuations reached when the channel is partially open, $G_0 < G_q$. We will 
find explicit analytic expressions for the correlation functions 
characterizing the evolution of the fluctuating differential capacitance with the magnetic field.                                     

We account for the Coulomb interaction by adding an extra term to the 
Hamiltonian: 
\begin{equation} \hat{H}=\hat{H}_0+E_C(\hat{n}-{\cal N})^2\;. 
\label{H} \end{equation} 
Here $\hat{H}_0$ is the Hamiltonian of the dot-lead 
system without Coulomb interaction, and the term $E_C(\hat{n}-{\cal N})^2$ 
accounts for the charging energy of the dot, $\hat{n}=\sum_p c_p^\dagger 
c_p^{\phantom{\dagger}}$ is the total number of electrons in the dot. The gate 
voltage $V_g$ is represented by the conventional dimensionless parameter ${\cal 
N}=V_gC_g/e$, with $C_g$ being the gate capacitance. The differential 
capacitance of the dot $C_{\it diff}$ is related to the ground state energy 
$E_0$ of the Hamiltonian (\ref{H}) by 
\begin{equation} C_{\it diff}({\cal N}) 
\equiv\frac{C}{C_g}\frac{\partial Q}{\partial V_g} 
=C\left(1-\frac{1}{2E_C}\frac{\partial^2 E_0}{\partial {\cal N}^2}\right)\;.
\label{QE0} 
\end{equation} 
Here $Q$ is the dot charge, $e\hat{n}$, averaged over 
the ground state of the system, $C$ is the capacitance of the isolated dot; we 
have introduced a lever-arm coefficient $C/C_g$ into the definition of $C_{\it 
diff}$.

The strength of tunneling through the dot-lead junction is 
characterized by its dimensionless conductance $g\equiv G_0/G_q$. To demonstrate 
the nonmonotonous behavior of the amplitude of the differential capacitance 
fluctuations, we consider analytically the limits of weak tunneling ($g\ll 1$) 
and of almost ballistic propagation of electrons in the junction ($1-g\ll 1$).

{\em Weak tunneling.}
The case of the weak tunneling, $g\ll 1$, is adequately described in the
tunneling Hamiltonian formalism,
\begin{eqnarray}
\hat{H}_0&=&
\sum_k \xi_k^{\phantom{\dagger}} a_k^\dagger a_k^{\phantom{\dagger}}
+\sum_p \xi_p^{\phantom{\dagger}} c_p^\dagger c_p^{\phantom{\dagger}}\label{Ht}\nonumber\\
&+&\sum_{k,p}(t_{kp} a_k^\dagger c_p^{\phantom{\dagger}}
+t_{kp}^*c_p^\dagger a_k^{\phantom{\dagger}})\;,
\end{eqnarray}        
where $\xi_k$ and $\xi_p$ are the one-electron energy spectra of the lead 
and dot respectively. Tunneling matrix elements here are related to $g$ by 
$g=\nu_l\nu_d\langle|t_{kp}|^2\rangle/2\pi^2$. Hereafter $\langle...\rangle$ 
denotes  the ensemble averaging, and $\nu_l$ and $\nu_d$ are the averaged 
densities of states in the lead and dot respectively. In the extreme $g\to 0$, 
the charge of the dot in the ground state is integer, $Q=me$. The half-integer 
values of the dimensionless voltage ${\cal N}$ correspond to the degeneracy 
points of the charging energy  in Eq.~(\ref{H}). If $V_g$ is tuned away from a 
degeneracy point ($m-1/2<{\cal N}<m+1/2$), it takes a finite energy $2E_C({\cal 
N}-m\pm 1/2)$ to increase or decrease the number of electrons in the dot by one 
with respect to the ground state.

The perturbation theory in the tunneling Hamiltonian (\ref{Ht}) can be used to
calculate the dot capacitance (\ref{QE0}), if the system is not too close to one of the
charge degeneracy points \cite{GlazmanMatveev}:
\begin{equation}
g^{\frac12} \ln(1/{\cal U})\ll 1\;.
\label{ptCondition}
\end{equation}
Here ${\cal U}\equiv\min\{|{\cal N}-m\pm1/2|\}$. The lowest-order correction to the ground state
energy is
\begin{equation}
E_{\it osc}({\cal N})=\frac{i}{2\pi}\frac{g}{\nu_{l0}\nu_{d0}}
\left.\int_{-\infty}^\infty dt\, {\cal G}_d(t;{\cal N}){\cal G}_l(-t)\right|_{{\bf r}
={\bf R}_c}\;,
\label{GreenE0}
\end{equation}
where ${\cal G}_d$ and ${\cal G}_l$ are the one-electron Green functions in the 
dot and the lead, determined by the Hamiltonians $\hat{H}_d$ and $\hat{H}_l$ 
respectively, and ${\bf R}_c$ is the point of the dot-lead contact. The density 
of states per unit area in the dot and in the lead are denoted here as 
$\nu_{d0}$ and $\nu_{l0}$. The dependence of ${\cal G}_d$ on the gate voltage is 
given by
\begin{eqnarray}
{\cal G}_d(t,{\cal N})&=&{\cal G}_{d0}(t)\left[\theta(t)e^{-2i(m+1/2-{\cal N})E_Ct}\right.\nonumber\\
&+&\left.\theta(-t)e^{2i({\cal N}-m+1/2)E_Ct}\right]\;,
\label{Gd}
\end{eqnarray}
where ${\cal G}_{d0}(t)$ is the one-electron Green function in the dot
determined by the Hamiltonian 
$\hat{H}_{d0}=\sum_p\xi_p^{\phantom{\dagger}} c_p^\dagger c_p^{\phantom{\dagger}}$.      
The exponential factors account for the change in the Coulomb energy of the system
when an extra electron or hole tunnels into the dot.

It follows from Eqs. (\ref{QE0}), (\ref{GreenE0}), and (\ref{Gd}) that the 
fluctuations of the differential capacitance $\delta C_{\it diff}$ are caused by 
the fluctuations of the one-electron Green function $\delta {\cal 
G}_{d0}\equiv{\cal G}_{d0}-\langle{\cal G}_{d0}\rangle$ in the dot at the point 
of the contact.   Then the statistical properties of $\delta C_{\it diff}$ can 
be related to  the correlation function  $\langle \delta{\cal G}_d(t_1;B_1)
\delta{\cal G}_d(t_2;B_2)\rangle|_{{\bf R}_c}$. The applied 
magnetic field $B$ affects the Green function ${\cal G}_{d0}$ thus affecting the 
capacitance fluctuations.

For a conducting dot, the correlation function for $\delta{\cal G}_{d0}$ is 
conventionally expressed in terms of the classical probabilities, diffuson 
${\cal D}$ and Cooperon ${\cal C}$\cite{AltshulerAronovReview}:
\begin{eqnarray}
&&\left.\langle \delta{\cal G}_{d0}(\omega_1;B_1)\delta{\cal
G}_{d0}(\omega_2;B_2)\rangle\right|_{{\bf R}_c} =2\pi \nu_{d0}\theta(-\omega_1\omega_2)\nonumber\\  &&
\times 
\left.\left[{\cal D}^{B_1,B_2}(|\omega_1-\omega_2|)
              +{\cal C}^{B_1,B_2}(|\omega_1-\omega_2|)\right]\;\right|_{{\bf R}_c}.
\label{wwDC}
\end{eqnarray}                             
For times exceeding the time of electron propagation across the dot of area $S$ 
or, equivalently, at energies much less than the Thouless energy for a closed 
dot, $E_T$, the probability density of an electron is homogeneously distributed over the dot, and
\begin{equation}
{\cal D}^{B_1,B_2}(\omega)=\frac{S^{-1}}{-i\omega+\Omega_-},\;
{\cal C}^{B_1,B_2}(t)=\frac{S^{-1}}{-i\omega+\Omega_+}.\;
\label{DCdot}
\end{equation}
Here $\Omega_\pm=E_T\left(S(B_1\pm B_2)/\Phi_0\right)^2$, and  $\Phi_0=2\pi\hbar 
c/e$ is the flux quantum. Differentiating Eq.~(\ref{GreenE0}) twice over ${\cal 
N}$ and extracting the random part, we come to the expression that relates 
fluctuations of the dot differential capacitance $\delta C_{\it diff}$ to the 
fluctuations of the one-electron Green function in the dot $\delta{\cal G}_{d}$. 
The correlation functions for  $\delta C_{\it diff}$  are thus related to those 
for $\delta{\cal G}_{d}$, which are given by Eqs. (\ref{Gd})--(\ref{DCdot}). 

First we consider  the variation of the  dot capacitance with the magnetic field 
at the same gate voltage. The corresponding correlation function is
\begin{eqnarray}
\langle \delta C_{\it diff}({\cal N},B_1)&&\delta C_{\it diff}({\cal N},B_2)\rangle\nonumber\\
&&=\frac{C^2g^2\Delta}{24\pi^4E_C{\cal U}^3}
\sum_\pm   M\left(\frac{B_\pm}{B_c({\cal U})}\right)
\label{dQdQB}
\end{eqnarray}
with $B_\pm=|B_1\pm B_2|$, ${\cal U}\equiv\min\{|{\cal N}-m\pm1/2|\}$, and
\begin{displaymath}
M(x)=\frac{6}{\pi}\int_0^\infty dy\;
\frac{x^2}{y^2+x^4}\;\frac{1}{y^3}\left[\frac{y^2+2y}{y+1}-2\ln(1+y)\right]\;.
\end{displaymath}
The function $M(x)\propto 1/x^2$ at large values of the argument $x\gg 1$, so 
that the correlation function decays with the increase of $B_\pm$ over the value 
$B_c$. At zero magnetic field,  the variance of the capacitance is 
\begin{displaymath}
\sqrt{\langle \delta C_{\it diff}^2\rangle} = \frac{1}{2\sqrt{3}\pi^2}
Cg\sqrt{\frac{\Delta}{E_C}}\,{\frac{1}{{\cal U}^\frac32}}\;.
\end{displaymath}
       
The characteristic scale for the variation of the correlation function 
Eq.~(\ref{dQdQB}) with the magnetic field is\cite{Bc}
\begin{equation}
B_c=\frac{\Phi_0}{S}\sqrt{\frac{U}{E_T}}\;.
\label{eBc}
\end{equation}
The meaning of this formula for the correlation field $B_c$ can be clarified if 
we recall the relation of the charge fluctuations to the electron Green 
function. In the semiclassical approximation, ${\cal G}(t)\propto e^{i{\cal 
S}(t)}$, where ${\cal S}$ is the action calculated along the trajectory. The 
dwelling time of an extra  electron in the dot is $t_d\sim\hbar/U$. Over this 
time, the trajectory of a diffusing electron would cover an area $Dt_d\sim \hbar 
D/U$, if there were no bounds imposed by the dot size. The ratio $n_d=\hbar 
D/SU\simeq E_T/U$ determines the typical number of ``winds'' for the electron 
trajectory inside the dot. Because of the random direction of the winds, the 
total flux threading the electron trajectory is $\Phi=BS\sqrt {n_d}$, and the 
correction to the action due to magnetic field is ${\cal S}_B\cong 
2\pi\Phi/\Phi_0$. Equating ${\cal S}_B$ to $2\pi$, we obtain the formula for 
$B_c$. 

Relatively short excitation lifetime, $t_d\ll\hbar/\Delta$, leads to 
correlations of the capacitance fluctuations in  different Coulomb blockade 
valleys. In a valley,  $\sim \hbar/\Delta t_d$ levels of spatial quantization 
closest to the Fermi level in the dot contribute to the charge fluctuations. 
Each subsequent peak of the Coulomb blockade corresponds to one more level 
occupied. Therefore, the fluctuations of capacitance at different valleys are 
correlated, unless the distance $n$ between them exceeds $E_C/\Delta$, so that 
an entirely new set of discrete levels contributes to the fluctuations. 
Calculation of the corresponding correlation function yields:     
\begin{equation}
\langle\delta C_{\it diff}({\cal U})\delta C_{\it diff}({\cal U}+n)\rangle\sim
\frac{C^2g^2\Delta}{4\pi^4E_C{\cal U}^3}\:
\frac{\ln\left(1+\displaystyle\frac{n\Delta}{2E_C{\cal U}}\right)}
        {\displaystyle\frac{\mathstrut n\Delta}{E_C{\cal U}}}\;.
\label{dQdQDifferentPeaks}
\end{equation}                                                                            

It is to be mentioned that Eqs.~(\ref{GreenE0}) and (\ref{dQdQB}) are valid only 
if condition (\ref{ptCondition}) is satisfied. If ${\cal N}$ is close enough to 
one of the charge degeneracy points, so that (\ref{ptCondition}) is violated 
(but ${\cal U}$ is still much greater than $\Delta/E_C$), the finite-order 
perturbation theory is insufficient for  calculation of the dot capacitance. 
Utilizing the renormalization group technique \cite{GlazmanMatveev}, we found 
that the capacitance fluctuations in this case are described by the same Eq. 
(\ref{dQdQB}) with the renormalized junction conductance
\begin{equation}
\tilde{g}=g\left[\cos^{2}\left(\frac{\sqrt{2g}}{\pi}\ln \displaystyle\frac{1}{\cal U}\right)\right]^{-1}\;.
\label{gtilde}
\end{equation}                                                                             
With this modification, the condition for the applicability of Eq.~(\ref{dQdQB}) 
is relaxed from (\ref{ptCondition}) to a weaker one, $g^{\frac12} \ln (1/{\cal 
U})\lesssim 1$. This new condition is imposed by the requirement that the 
renormalized conductance $\tilde{g}$ must be still much less than unity.

{\em Strong tunneling}. As it was pointed out in the introduction, the effects of the
Coulomb blockade become less pronounced as the dimensionless conductance of the
dot-lead junction, $g$, approaches unity. In the limit $\Delta=0$,
Flensberg\cite{Flensberg} and Matveev\cite{Matveev} studied  the case of  a dot
connected to the lead by a constriction with small reflection amplitude
$r=\sqrt{1-g}$. Since the characteristic energy $E_C$ is much smaller than the Fermi
energy, one can linearize the spectrum of one-dimensional electrons in the
constriction.  The electrons in the constriction are thus divided into left- and
right-moving species, $\psi(x)=e^{-ik_Fx}\psi_L(x)+e^{ik_Fx}\psi_R(x)$.
In\cite{Matveev}, the full Hamiltonian (\ref{H}) describing the electrons in the dot
and lead connected to each other, is transformed to an effective one-dimensional form
and expressed in terms
$\psi_L$, $\psi_R$. In the limit $\Delta\to 0$ and in the absence of backscattering in
the constriction, there is no mixing between the $L$ and $R$ states:
\begin{eqnarray}
\hat{H}_{\rm
1D}&=&iv_F\int_{-\infty}^{\infty}dx\left\{\psi^\dagger_L\partial_x \psi^{\phantom{\dagger}}_L-
\psi^\dagger_R\partial_x\psi^{\phantom{\dagger}}_R\right\}\nonumber\\
&+&E_C\left(\int_{-\infty}^0dx:\psi^\dagger_L\psi^{\phantom{\dagger}}_L+
\psi^\dagger_R\psi^{\phantom{\dagger}}_R:-{\cal N}\right)^2;\label{1DH}
\end{eqnarray}
here $v_F$ is the Fermi velocity, and $: \ldots :$ stands for the normal ordering. The two terms
in the Hamiltonian (\ref{1DH}) account for the kinetic energy of the electrons and charging
energy, respectively.  Both scattering from a barrier in the
constriction, and scattering from within a finite-size dot, create terms
$\propto (\psi_R^\dagger\psi_L^{\phantom{\dagger}}+h.c.)$ in the Hamiltonian, and in the
corresponding action \cite{AleinerGlazman97}. The backscattering terms lead to
essentially non-perturbative corrections to the ground state energy
$E_0$. In the work of Matveev \cite{Matveev},  the problem with a single scatterer in the
constriction described by
\begin{equation}
\hat{H}_{bs}=v_F\left.\left(r\psi^\dagger_L\psi^{\phantom{\dagger}}_R+
r^*\psi^\dagger_R\psi^{\phantom{\dagger}}_L\right)\right|_{x=0}\;,
\label{Hbs}
\end{equation}
was solved, and $E_0$ was found. According to\cite{Matveev}, the regular Coulomb blockade
oscillations in the energy are given by $E_{\it osc}({\cal N})\sim 2E_C|r|^2\cos (2\pi{\cal
N})\ln\left[1/|r|^2\cos^2(\pi{\cal N})\right]$ at
$r\neq 0$.

Backscattering from inside the cavity ($\Delta\neq 0$) generates a random contribution to
$E_0$, superimposed on the regular oscillations. Once the reflection in the channel is
accounted for non-perturbatively,  the random backscattering from the walls of the dot can be
studied by means of the lowest-order perturbation theory \cite{AleinerGlazman97}:
\begin{equation}
\delta E_0=i\frac{v_F^3}{\nu_{l0}}\int_{-\infty}^{\infty} dt\,\left.\delta{\cal G}_{d0}(t)
G_{LR}(-t;{\cal N})\right|_{{\bf r}={\bf R}_c}\;,
\label{dOmegaTime}
\end{equation}
 where $G_{LR}$
is a component of the electron Green
function in a 1D infinite wire, described by the Hamiltonian
$\hat{H}=\hat{H}_{1D}+\hat{H}_{bs}$ [see Eqs.~(\ref{1DH}), (\ref{Hbs})],
\begin{equation}
G_{LR}(t)=\langle\psi^\dagger_R(t)\psi^{\phantom{\dagger}}_L(0)
+\psi^\dagger_L(t)\psi^{\phantom{\dagger}}_R(0)\rangle|_{x=0}\;.
\label{GLR}
\end{equation}
Function $G_{LR}$ describes the dynamics of backscattering of an individual
electron  from the dot. An electron wave in the channel carries charge and spin
degrees of freedom, and there are two time scales involved in the backscattering process. The
shorter one,
$t_\rho\sim\hbar/E_C$, corresponds to the reflection of the charge by the Coulomb barrier
(resulting from the charging energy). The longer one, $t_\sigma$, describes the reflection of the
spin component, and at finite  $r$ is of the order of $1/E_{\it osc}({\cal N})$. 
At the intermediate time scale, $1/E_C<|t|<t_\sigma$, the charge
and spin components of the
backscattered electron are separated, and the Green function manifests a
distinctively
non-Fermi-liquid behavior,
\begin{equation}
G_{LR}(t;{\cal N})=
\frac{\cos\pi{\cal N}}{\pi v_F t}
\sqrt{
\displaystyle\frac{\gamma\mathstrut}{\pi^3\mathstrut}
\displaystyle\frac{|t|}{t_\sigma({\cal N})}
}
\ln\left(\displaystyle\frac{t_\sigma({\cal N})}{|t|}\right)\;.
\label{G2}
\end{equation}
Here  $\gamma=e^{\bf C}$ with ${\bf C}=0.577$ being the Euler constant. The electron
backscattering process is completed at $t\sim t_\sigma$, and at
larger time scales the system exhibits the standard Fermi-liquid behavior with
$G_{LR}(t)\sim 1/t$. The result (\ref{G2}) can be obtained in the first-order 
perturbation theory in $\hat{H}_{bs}$, with the low-energy cut-off $E\sim E_{\it
osc}({\cal N})$.

Equations (\ref{QE0}) and (\ref{dOmegaTime}) enable us to relate capacitance 
fluctuations $\delta C_{\it diff}$ to one-electron properties of the system 
described by the functions $\delta {\cal G}_{d0}$ and $G_{LR}$.  The major 
contribution to the integral (\ref{dOmegaTime}) is given by the interval 
$1/E_C<|t|<t_\sigma$, where approximation (\ref{G2}) is applicable. The 
correlation function of the capacitance fluctuations is then calculated with the 
help of Eqs.~(\ref{wwDC}) and (\ref{DCdot}):
\begin{eqnarray}
&&\langle\delta C_{\it diff}({\cal N}, B_1)\delta C_{\it diff}({\cal M}, B_2)\rangle=
\frac{16\gamma}{3\pi^2}\,C^2\frac{\Delta}{E_C}|r|^2\nonumber\\
&&\times\cos2\pi{\cal N}\cos2\pi{\cal M}
\ln\left(\frac{1}{|r|^2\cos^2\pi{\cal N}}\right)
\ln\left(\frac{1}{|r|^2\cos^2\pi{\cal M}}\right)\nonumber\\
&&\times\sum_\pm P\left(\frac{B_\pm^2}{B_c^2};r^2_{\it max}\right)\;,\;\;|r|^2\equiv 1-g\;,
\label{phaqit}
\end{eqnarray}
with $r_{\it max}\equiv\max\{|r\cos\pi{\cal N}|,|r\cos\pi{\cal M}|\}$ and
$B_c\equiv(\Phi_0/S)\sqrt{2E_C/E_T}$. 
 The variation of the correlation function with the magnetic
field is described by the function $P(B^2/B_c^2,r^2_{max})$. In the domain $B<B_c$,
\begin{displaymath}
P\left(\frac{B^2}{B_c^2};r^2_{\it max}\right)\simeq
\left[\ln\displaystyle\frac{1}{r^2_{\it max}}\right]^3
-\left[\ln\max\left\{1;\frac{B^2}{r^2_{\it max}B^2_c}\right\}\right]^3\;.
\end{displaymath}
At magnetic fields larger than the correlation field $B_c$, the correlations rapidly
decay, $P\left(B^2/B_c^2;r^2_{\it max}\right)\propto (B_c/B)^4$.

The perturbation theory Eq.~(\ref{dOmegaTime}) is applicable if the channel 
conductance is not too close to $G_0$, {\it i.e.}, $|r|^2\gg\Delta/E_C$. At a 
larger conductance, the amplitude of fluctuations saturates at $\delta 
C\sim C(\Delta/E_C)\ln^2(E_C/\Delta)$ \cite{AleinerGlazman97}.

{\em Tunneling of arbitrary strength.} 
The behavior of the capacitance fluctuations in the whole domain $0\leq G_0\leq 
G_q$ (with $G_q=e^2/\pi\hbar$) follows from Eqs. (\ref{dQdQB}) and 
(\ref{phaqit}). At $G_0\ll G_q$ the fluctuations increase with conductance, 
$\delta C_{\it diff}/C\sim (G_0/G_q)\sqrt{\Delta/E_C}$.  For an almost open 
channel ($G_q-G_0\ll G_q$) the characteristic amplitude of the capacitance 
fluctuations is $\delta C_{\it diff}/C\sim [1-(G_0/G_q)]\sqrt{\Delta/E_C}$, up to a 
$\ln^3$ factor. Thus the mesoscopic capacitance fluctuations reach maximum at a 
partial transmission of the channel connecting dot and lead. The parameter 
$\sqrt{\Delta/E_C}$ for typical semiconductor dots \cite{Kouwenhoven} is $\sim 
0.2$, so that the relative magnitude of these fluctuations is appreciable.

In conclusion,  we found mesoscopic fluctuations of charge and differential 
capacitance of a partially opened quantum dot. The amplitude of fluctuations 
reaches maximum at the intermediate junction conductance, $G_0\lesssim 
e^2/\pi\hbar$. We  also found the correlation function of these fluctuations at 
different magnetic fields and gate voltages. In the experiments 
\cite{Westervelt,Molenkamp,Zhitenev} no special attention was paid to the 
mesoscopic fluctuations. However, the parameters of quantum dots in 
\cite{Westervelt,Molenkamp,Zhitenev} are in the right domain for studying 
mesoscopic fluctuations of the charge and of the differential capacitance. This 
can be accomplished by  observing the Coulomb blockade with a magnetic field 
applied, which causes random variations of the charge.

The work at the University of Minnesota was supported by NSF Grants DMR-9423244 and DMR-9731756.

\end{multicols}
\end{document}